# A Formal Approach to Distributed System Tests Design

Andrey A. Shchurov
Department of Cybernetics
Czech Technical University in Prague
Prague, the Czech Republic
E-mail: shchuand {at} fel.cvut.cz

Radek Mařík
Department of Telecommunications Engineering
Czech Technical University in Prague
Prague, the Czech Republic

*Abstract*—**Deployment of distributed systems sets high requirements for procedures and tools for the complex testing of these systems. This work introduces a formal four-layered model for test generation mission on the basis of the component-based approach and the concept of layered networks. Based on this model, we describe the test generation strategy that covers every interaction from the end-user requirements on all coexisting architectural layers, and checks the internal consistency of the system technical specifications with respect to the end-user requirements. The next step introduces the Prolog-based approach to representing this model and the requirements-coverage strategy**

## I. INTRODUCTION

*You know you have a distributed system when the crash of a computer you've never heard of stops you from getting any work done.*

*- Leslie Lamport*

This maxim accurately reflects the nature of distributed systems. For the engineering world a more formal definition was given by Andrew S. Tanenbaum and Maarten van Steen as *a collection of independent computers that appears to its users as a single coherent system* [1]. But in practice, this definition can denote:

- a collection of components/products (hardware and software) – the viewpoint of the vendor community;
- a collection of the above plus external communication infrastructure – the viewpoint of the network engineer community;
- a collection of services/applications – the viewpoint of the software/system engineer community;
- all of the above plus end-users/customers – the viewpoint of the business community.

When talking about the testing of distributed systems, the confusion between these definitions is a fertile source of vulnerabilities. Broadly speaking, vendors focus on individual component testing problems only – but in general testing or qualification of the elements of the system does not cover the system testing itself. In turn, network engineers usually focus on network subsystem testing. In this case, ignoring services/applications testing is one of the most common causes of system problems [2]:

- If the network subsystem is not solid, services/applications cannot be responsive and reliable by definition.
- If the network subsystem is solid, but the services/applications do not provide required performance or functionality, end-users could perceive the network subsystem as unavailable or unreliable.

On the other hand, distributed systems differ from traditional software because components are dispersed across a network. Very often software/system engineers do not take this dispersion into account that leads to the following false assumptions about the network subsystem: (1) networks are always reliable; (2) latency is zero; (3) bandwidth is infinite; etc. [1]. As a consequence, only the business (or integration) viewpoint brings all of the detailed elements of the distributed systems together through a process of testing (or qualification) to achieve a valid system for meeting the ultimate needs of the end-users [3].

In addition, virtualization and cloud technologies set another level of system complexity. In this case, there is a "classical" physical (external) system as a host system and some (at least one) virtual (internal) systems. It is a real challenge for testing activities and/or testing applications. Even if we could test a host system and every virtual system independently, the amount of work would increase several times.

Our main goal is the automated design and generation of testing procedures/specifications and plans for distributed systems based on end-user requirements and technical specifications as a necessary part of project documentation. Working engineers treat formal methods as they are widely taught in universities and not used anywhere in the real world. But in the case of complex or non-standard systems, personal experience and/or intuition are often inadequate. Thus, to accomplish such a goal we need to identify a top level test philosophy (or test strategy) based on a formal model with the following criteria: (1) it must be appropriate for the viewpoint of the business community; (2) it should be based on standards (as a formal document); (3) it has to cover all aspects of distributed systems; and (4) it has to be simple enough for practical application.

In this paper we come after the basic steps of the following general methodology of system/software test templates design. At first, we specify entities and their relations of the distributed system under test (DSUT). Both the entities and the tuples of relations are assigned unique abstract identifiers. Their attributes are determined by a set of facts about the abstract identifier expressed as atomic propositions. Type information on a given entity is also covered by these atomic propositions. Thus, the system under test is modeled as a weighted graph structure. Secondly, we need to specify test requirements. The test requirements can be considered as mappings between an input domain and an output space of test templates. The input domains consist of subsets of entities and their relations. A test template of the output space might be itself a complex structure that specifies all necessary conditions of the given test template. For example, in our case, test templates are often paths between given nodes of the graph model of the distributed system. Although a test template has its internal structure it can be treated as a single point in a multidimensional space of all test templates. Finally, the test template space is constructed through an application of test requirement mappings onto the input specification graph model. In this paper we deal exclusively with such a test templates space creation. Its optimization and convertion into test procedures and their integration into a test suite is beyond the scope of this paper.

The rest of this paper is structured as follows. Section 2 introduces the related work. Section 3 presents the formal multilayer model of distributed systems for test generation mission and the test templates generation strategy. Section 4 focuses on the Prolog-based approach to representing the model and the strategy. Section 5 discusses which features of distributed systems can be covered. Finally, conclusion remarks and future research directions are given in Section 6.

## II. RELATED WORK

The automated generation of test templates is the most important role for formal verification in testing. It involves analyzing system models, with the analysis covering paths in a model. In this context, this work lies in the area of model-based testing (MBT), but it differs from existing approaches in the way of deriving models. MTB research in the domain of distributed systems can be roughly classified into three categories [4], [5], [6]:

1. *MTB general approaches.* El-Far and Whittaker [7] give a general introduction to principle, process, and techniques of model-based testing. In turn, Stocks and Carrington [8] suggest that test templates can be defined as the base for test case generation and a large test template is divided into smaller templates for generating more detailed test cases.

2. *MBT based on graphical models.* Offutt and Abdurazik [9] describe an approach to generating test cases from UML Statecharts for components testing. And Hartmann et al [10] extend the approach for integration testing and for test automation.

3. *MBT based on formal specifications.* Bernot et al [11] set up a theoretical basis for specification-based testing, explaining how a formal specification can serve as a base for test case generation. Dick and Faivre [12] propose to transform formal specifications into a disjunctive normal form (DNF) and then use it as the basis for test case generation. Donat [13] represents a technique for automatically transformation formal specifications into test templates and taxonomy for coverage schemes. Hong et al [14] show how coverage criteria based on control-flow or data-flow properties can be specified as sets of temporal logic formulas, including state and transition coverage as well as criteria based on definition-use pairs.

The most recent systematic method is presented by Liu and Shen [15]. This method can be used for (1) identifying all interface scenarios, formalizing requirements into formal operation specifications whose interfaces are consistent with the corresponding ones of the program; and (2) for testing programs based upon the formal specifications (scenario-coverage strategy).

But in spite of tremendous efforts of many researchers, MBT is still difficult to be used for large scale systems due to its complexity and the potential inconsistencies in both component and system architectures. Furthermore, it still considers hardware-based and software-based systems independently and, as a consequence, the layered structure of modern communication protocols (such as OSI layered architecture [16]) is completely ignored.

## III. FORMAL MODEL OF DISTRIBUTED SYSTEMS

### A. Basic approach

The essential idea of our approach is based on two basic notions:

- component-based approach with its two important consequences: (1) components are built to be reused in different systems, and (2) component development process is separated from the system development process [17], [18];

- concept of layered complex networks [19].

The component-based approach refers to the fact that the functional usefulness of distributed systems do not depend on any particular part of these systems, but emerges from the way in which their components interact. Thus, the standard ISO/OSI Reference Model (RM) [16] can be used as a starting point. But it cannot cover the all required aspects by oneself (practically it is necessary to use several models in order to cover many different aspects). Necessary complements to RM provide the set of architectural models [20] as the most intuitive solution. In turn, the concept of layered complex networks secures the consistency between different models. And finally the graph theory (as a standard-de-facto) provides the necessary tools for models representations.

For our purposes the system model can be stated as a four-layered graph as follows:

- The ready-for-use system architecture layer defines functional components and their interconnections. This layer is based on functional models [20] (the enlarged viewpoint of end-users/customers) and covers the application (L7) layer of RM.

- The service architecture layer defines software-based components (services/applications) and their interconnections. It is founded on flow-based models [20] (representation of centralized or client-server, decentralized or peer-to-peer, and hybrid architectures) and covers the transport (L4), session (L5), presentation (L6) and partially application (L7) layers – we cannot divide these layers in the case of commercial off-the-shelf (COTS) software.

- The logical architecture layer defines logical (virtual) components and their interconnections. It is based on topological models [20] and covers the network (L3) layer.

- The physical architecture layer defines hardware (physical) components and their interconnections. Like its predecessor, this layer is founded on topological models but covers the physical (L1) and data link (L2) layers – we cannot divide these layers in the case of COTS telecommunication/network equipment.

- The interlayer projections define all types of components hierarchical (interlayer) relations/mapping. These relations make the layered model consistent and convenient for our goals.

In the real world neither owner nor developer can be able to test its distributed system completely (entire systems and every component in it) due to the lack of resources. Thus, what it should be done is development of a test philosophy (or test strategy) that covers critical aspects of distributed systems.

*B. Formal notations*

Formal verification offers a rich toolbox of mathematical techniques that can support the model-based testing of computer systems. This toolbox contains logic programming as one of the most relevant technique of model checking [5]. In turn, logic programming deals with logical facts and, as a consequence, the first step is to determine the formal notations which make the layered model applicable for logic programming.

**Definition 1 (the system model):** *Let the graph G denote DSUT:*

$$G = (V, E, M)$$

*where G is multi-layered 3D graph, derived from the DSUT specification; V(G) is a finite, non-empty set of components of DSUT; E(G) is a finite, non-empty set of component-to-component connections; and m(G) is a finite, non-empty set of component-to-component interlayer mapping (or projections). Then, the system model $G_n$ for each layer n can be represented as a subgraph of G:*

$$G_n = (V_n, E_n, M^n_{n-1}, V_{n-1})$$

*where $V_n(G_n)$ is a finite, non-empty set of components of DSUT on layer n; $E_n(G_n)$ is a finite, non-empty set of component-to-component connections on layer n; $M^n_{n-1}(G_n)$ is a finite set of component-to-component projection from layer n to layer n-1; and $V_{n-1}(G_n)$ is a finite set of components of DSUT on layer n-1.*

Generally, $G_n$ is intransitive by default with the exception of the physical architecture layer. In turn, $M^n_{n-1}(G_n)$ and $V_{n-1}(G_n)$ must be non-empty sets with the same exception – in this case, the definition of projection has no physical meaning.

The systems decomposition into objects which interact is a common baseline for all technologies for the design and implementation of distributed systems [21]. These two aspects (components and information links) of knowledge are usually included in formal specifications [22]. As a consequence, the test requirements for every layer can be derived from formal/technical specifications and can be formalized into a set of formal operations. We need to state here that the tests of individual components are usually prepared by vendors (see the viewpoint of the vendors' community). In the real world we have to rely on:

- vendors and/or independent laboratories information about products;

- vendors conclusions of products compliance with end-user requirements;

- vendors documentation (includes test descriptions).

**Definition 2 (the test requirements):** *Let the set R denote the requirements for DSUT specified by the end-users. For simplicity, R can be represented as:*

$$R = R_{comp} \cup R_{dist},$$

*where $R_{comp}$ is a finite set of individual components requirements; and $R_{dist}$ is a finite set of distributed aspects (component-to-component communication) requirements. In this case, the distributed aspect requirements $R_n$ for each layer n can be represented as a set of operators $R_{ni}$ (i = 1 … r) that are applied to the system model $G_n$ and produce test templates $T_n$ as its output, often as paths between given pairs of vertices:*

$$R_n = \{R_{n1}(s_{n1}, t_{n1}, c_{n1}),…, R_{nr}(s_{nr}, t_{nr}, c_{nr})\}$$

*where $s_{ni}$, $t_{ni}$ is a pair i of individual components on layer n, which must communicate; and $c_{ni}$ are technical characteristics of component-to-component communication processes (information links specifications).*

Generally, $R_n$ can be an empty set with the exception of the end-user (top-level) requirements (in this case, the definition of distributed systems makes no sense at all).

At this place to avoid confusions, we should stressed that the term of "test requirement" denotes in classical software testing publications both a process of searching for what should be tested and entities that are produced by such a process, e.g. in [23]. As the process is accomplished by human testers, using the same term for both items does not cause any harm. However, in this paper we deal with an automation of such process and we need to reference such items without any ambiguity. Therefore, we use the term "test requirement" to denote the process of searching, and "test template" to denote entities that are results of such a process. We do not use the term "test case" as we often receive a set of test templates that are only the bases for test cases design.

As a next step, before describing the test strategy, we need to clarify the formal concept of test templates.

**Definition 3 (the test templates):** *Let the set $T_n$ denote the test templates for each layer n as a set of pairs:*

$$T_n = \{(P_{n1}(s_{n1}, t_{n1}), c_{n1}), \ldots, (P_{nr}(s_{nr}, t_{nr}), c_{nr})\}$$

*where $P_{ni}(s_{ni}, t_{ni})$ is a path (or data flow) between $s_{ni}$ and $t_{ni}$ in $G_n$; and $c_{ni}$ are technical characteristics of component-to-component communication processes.*

The set of test templates $T_n$ cannot be an empty set.

In contrast to the multilayer model of complex systems [19], the sets of nodes of the system model (see Definition 1) on each layer are not identical. And the key factor is the arity of the component-to-component projection from layer *n* to layer *n-1* (the top-down point of view). This parameter allows representing technological solutions used to build the system:

- $1_n : N_{n-1}$, e.g. clustering/stacking technology;
- $N_n : 1_{n-1}$, e.g. virtualization/replication technology;
- $1_n : 1_{n-1}$, e.g. a special case of dedicated components.

**Definition 4 (the interlayer projection):** *Let the set of operators $I^n_{n-1}$ be the mapping of the test templates $T_n$ from layer n to layer n-1:*

$$R^n_{n-1} = I^n_{n-1}(G_n, T_n)$$

$R^n_{n-1}$ has the same structure as $R_n$ (see Definition 2) with the exception of parameter $c_{ni}$ – in the case of $R^n_{n-1}$, $c_{ni}$ can be empty:

$$R^n_{n-1} = \{R^n_{(n-1)1}(s_{(n-1)1}, t_{(n-1)1}, c_{(n-1)1}), \ldots\}$$

Generally, at least one requirement of $R^n_{n-1}$ must be mapped for each individual test template of $T_n$ because every path on layer *n* must have at least one projection on layer *n-1*. As a consequence, $R^n_{n-1}$ cannot be an empty set.

### C. Test strategy

A test strategy (or test philosophy) establishes *what* should be tested and *why* [2]. Based on the definition of the test requirements (see Definition 2), we can determine the two main steps of the test strategy:

**Individual components testing:** The first subset of test templates is the node covering of the system model *G* (tests that confirm that the nodes are there) [23].

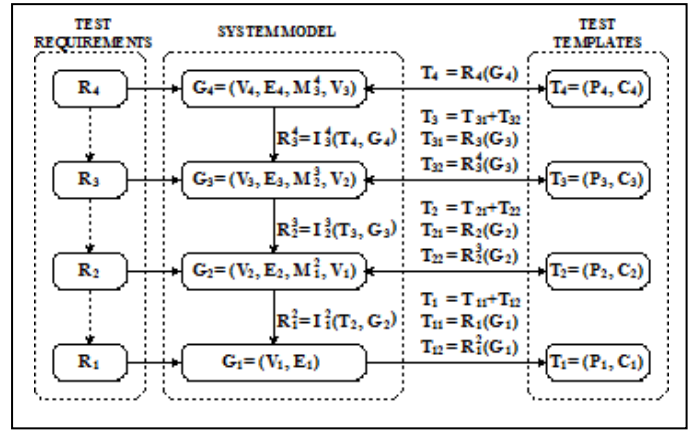

Figure 1. Graphical representation of the requirements-coverage strategy

This subset is the function of the system specification and it is based on the formalism for object representation.

**Distributed aspects testing:** As was mentioned before, DSUT has a hierarchical (layered) architecture. As a consequence, DSUT must satisfy two conditions for each layer *n*: (1) set of requirements for layer n; and (2) set of requirements, defined by upper layer projections (if this layer exists).

As a consequence:

- The first subset of test templates, denoted by $T_{n1}$, is the result of applying of the test requirements $R_n$ to the system model $G_n$ for each layer n:

$$T_{n1} = R_n(G_n)$$

**Criterion 1:** At least one test template of $T_{n1}$ must exist for every individual requirement of $R_n$. If $T_{n1}$ is an empty set, it means that the technical specification is internally inconsistent with respect to the end-user requirements (see Definitions 1 – 3).

- The second subset of test templates, denoted by $T_{n2}$, is the result of applying of the interlayer requirements $R^{n+1}_n$ to the system model $G_n$ for each layer n:

$$T_{n2} = R^{n+1}_n(G_n)$$

**Criterion 2:** At least one test template of $T_{n2}$ must exist for every individual requirement of $R^{n+1}_n$ (see Criterion 1 and Definition 4).

Then, the set of test templates $T_n$ for each layer *n* is a union of its subsets:

$$T_n = T_{n1} \cup T_{n2}$$

This requirements-coverage test strategy (see Fig. 1) states that for each requirement there exists at least one test template based on its conditions. If the system meets all these conditions, it will ensure that (1) every interaction from the functional system architecture layer based on end-user requirements is tested (at least once) on all coexisting, i.e. functional, system, logical and physical, architectural layers (see Criteria 1 and 2), and, therefore, (2) the technical specification is internally consistent with respect to the end-user requirements (see Definitions 2 and 4).

D. *Complexity of the approach*

The total number of test templates $T$ is based on the two main steps of the test strategy:

$$T = T_{comp} + T_{dist} \quad (1)$$

where $T_{comp}$ is the number of individual components test templates; and $T_{dist}$ is the number of distributed aspect (information links) test templates.

In the case of components test templates the result is a quite trivial:

$$T_{comp} = C = \sum_{n=1}^{N} C_n \quad (2)$$

where $N$ is the number of DSUT layers; $C$ is the total number of DSUT individual components; and $C_n$ is the number of individual components on layer $n$.

As was mentioned before, each distributed aspect test template defines as a path between a given pair of DSUT individual components (see Definition 3). As a consequence, the set of test templates for each layer $n$ can be represented as an induced subgraph of $G_n$:

$$G'_n \subseteq G_n, \ s_{ni} \in V(G'_n), t_{ni} \in V(G'_n) \quad (3)$$

where $s_{ni}, t_{ni}$ is a pair $i$ of individual components on layer $n$, which must communicate.

The maximum possible number of test templates (the maximum possible number of subgraph edges) can be achieved if the induced subgraph $G'_n$ is a complete graph [24]. There are two possible options:

1. *Simple communication systems*. In this case, there is only one possible route between each pair of individual components, which must communicate. For such systems:

$$T_{dist} \leq \sum_{n=1}^{N} \frac{|G'_n|(|G'_n| - 1)}{2} \quad (4)$$

And:

$$T \leq \sum_{n=1}^{N} \left[ C_n + \frac{|G'_n|(|G'_n| - 1)}{2} \right] \quad (5)$$

2. *Complex communication systems*. For these systems there are some independent routes between each pair of individual components, which must communicate. In this case:

$$T_{dist} \leq \frac{1}{2} \sum_{n=1}^{N} \sum_{i=1}^{|G'_n|} \sum_{j=1}^{|G'_n|} r_{nij}, \ r_{nij} = \begin{cases} r_{nij} = r_{nji}, i \neq j \\ 0, \quad i = j \end{cases} \quad (6)$$

where $r_{nij}$ is the number of parallel/redundant paths for each pair of individual components on layer $n$, which must communicate. In turn:

$$T \leq \sum_{n=1}^{N} \left[ C_n + \frac{1}{2} \sum_{i=1}^{|G'_n|} \sum_{j=1}^{|G'_n|} r_{nij} \right] \quad (7)$$

The next step is based on the following assumptions:

- The number of DSUT layers is limited by the system model (see Definition 1): $N = 4$.
- In the real engineering word under financial constraints commercial systems are usually based on redundant architecture [25]: $r_{nij} = 2$ (specific areas like the military, nuclear or aerospace industries are beyond the scope of this work).
- All individual components on each layer must communicate: $|G'_n| = |G_n| = C_n$.

In the case of simple communication systems:

$$T \leq \sum_{n=1}^{4} \left[ C_n + \frac{C_n(C_n - 1)}{2} \right] = \sum_{n=1}^{4} \frac{C_n(C_n + 1)}{2} \quad (8)$$

In turn, in the case of complex communication systems:

$$T \leq \sum_{n=1}^{4} \left[ C_n + \frac{2C_n(C_n - 1)}{2} \right] = \sum_{n=1}^{4} C_n^2 \quad (9)$$

As a consequence, the relation between the DSUT size and the number of test templates in the case of commercial systems can be represented as a quadratic dependence.

IV. PROLOG-BASED APPROACH

As a programming language, Prolog is especially well suited for problems that involve objects and relations between them [26]. It marks Prolog as the most relevant tool in our case (see Definitions 1 – 4). Based on these formal notations, we can consider three main building blocks: (1) system model; (2) test requirements; and (3) test strategy. In the following three subsections we deal with a brief overview of these Prolog-based blocks and their examples.

*A. System model representation*

The system model itself can be represented as the following Prolog facts (every individual element is represented as a single fact).

1. Facts for components representation (multi-layered graph nodes):

**object_(layer_(N), component_(Class, Num), type_(Type), parameter_([Param])).**

where *layer_(N)* is a layer identifier (*N* = 1 – physical architectural layer, *N* = 2 – logical architectural layer, *N* = 3 – system architectural layer, *N* = 4 – functional architectural layer); *component_(Class, Num)* is a component identifier (the nodes of the multi-layered graph); *type_(Type)* is a component descriptions; and *[Param]* is a list of component technical parameters.

The variable *Type* can be defined as an empty list (see example below) in the case of objects that cannot be associated with hardware or software components (virtual objects on layer 2 and layer 4). The list *Param* is a configuration of hardware-based components on layer 1 and hardware requirements for software-based components on layers 2 and 3. This information must be used for control of appropriate distribution of hardware resources (specification control) across the distributed system.

2. Facts for component-to-component connections representation (multi-layered graph edges):

**connection_(layer_(N), component_(Class_1, Num_1), component_(Class_2, Num_2), parameter_([Param])).**

where *component_(Class_1,Num_1)* and *component_(Class_2, Num_2)* is a pair of interacting components on layer *N* (the edges of the multi-layered graph); and *[Param]* is a list of technical parameters of component-to-component connections (ports, protocols, modes, etc.).

Generally, the system model is an undirected graph. Thus, the facts *connection_()* must be supplemented by the predicate *relation_()*:

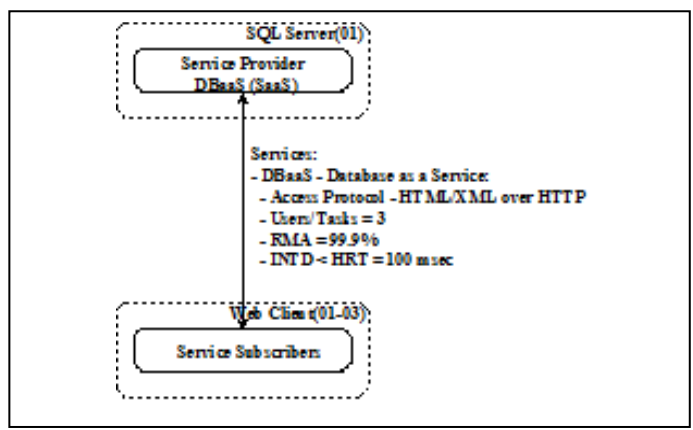

Figure 2. Simple example of multy-layered system model – functional architecture layer

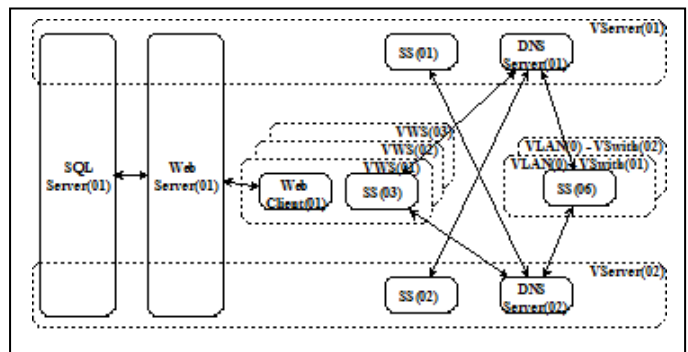

Figure 3. Simple example of multy-layered system model – service architecture layer

**relation_(N, [X1, X2], [Y1, Y2]) :-
connection_(layer_(N), component_(X1, X2), component_(Y1, Y2), _)
;
connection_(layer_(N), component_(Y1, Y2), component_(X1, X2), _).**

3. Facts for component-to-component interlayer projections representation (multi-layered graph edges):

**map_(layer_(N), component_(Class_1, Num_1), component_(Class_2, Num_2), parameter_([Param])).**

where *component_(Class_1,Num_1)* and *component_(Class_2, Num_2)* is an interlayer projection from layer *N* to layer *N-1*; and *[Param]* is a list of technical parameters of component-to-component interlayer projection (clustering, virtualization, etc.).

With the exception of the bottom layer 1, every component must have at least one top-down projection, BUT it is not necessary for every component to have a bottom-up projection (for example – system services are always hidden from the end-user viewpoint).

Based on the representation defined above, the simple example of multi-layered system model (see Fig. 2 – Fig. 5) can be represented as a list of the Prolog predicates (some lines are skipped and all lists of technical parameters such as hardware configurations, communication protocols, ports, etc. are shown empty because of the lack of space):

**object_(layer(4), component_(provider,1),
type_([]), parameters_([])).
object_(layer(4), component_(subscriber,1),
type_([]), parameters_([])).
object_(layer(3), component_(sql_server,1),
type_('MySQL Server 5.6'), parameters_([])).
object_(layer(3), component_(web_server,1),
type_('Apache 2'), parameters_([])).
object_(layer(3), component_(web_client,1),
type_('Firefox 29.0.1'), parameters_([])).**

**object_(layer(3), component_(web_client,2),
type_('Firefox 29.0.1'), parameters_([])).**

**connection_(layer(4), component_(provider,1),
component_(subscriber,1), parameters_([])).**
**connection_(layer(3), component_(web_server,1),
component_(sql_server,1), parameters_([])).**
**connection_(layer(3),component_(web_server,1),
component_(web_client,1), parameters_([])).**
**connection_(layer(3),component_(web_server,1),
component_(web_client,2), parameters_([])).**

**map_(layer(4),component_(provider,1),
component_(sql_server,1), parameters_([])).**
**map_(layer(4),component_(subscriber,1),
component_(web_client,1), parameters_([])).**
**map_(layer(4),component_(subscriber,1),
component_(web_client,2), parameters_([])).**

An important note: on layer 3 (system architectural layer) we have to separate application and system services on independent subsystems not directly (via facts *connection_()*) related to each other. The maxim is always "do not add more detail than is necessary". It is essential to avoid "phantom" paths in the system model which can be created based on formal representation only. For example, two facts:

**connection_(layer(3), component_(web_server,1),
component_(ss,1), parameters_([])).**
**connection_(layer(3), component_(web_client,1),
component_(ss,3), parameters_([])).**

can cause a "phantom" path *[web_server, 1] ↔ [ss, 1] ↔ [dns_server, 1] ↔ [ss, 3] ↔ [web_client, 1]*. In this case, relations between components can be set via common interlayer projections, e.g. the relation between *component_(web_server, 1)* and *component_(ss, 1)* is defined by the pair of projections to common *component_(vserver, 1)*:

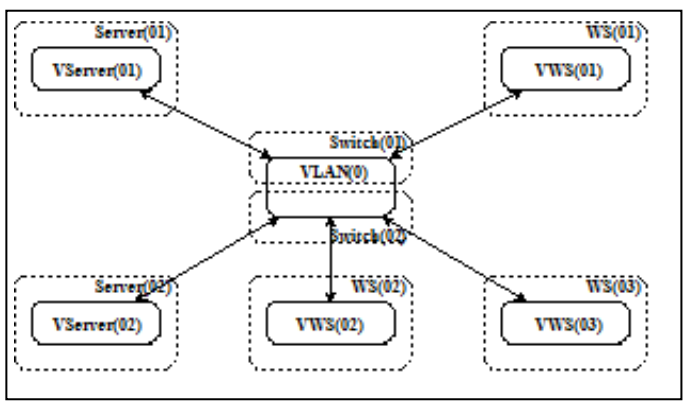

Figure 4. Simple example of multy-layered system model – logical architecture layer

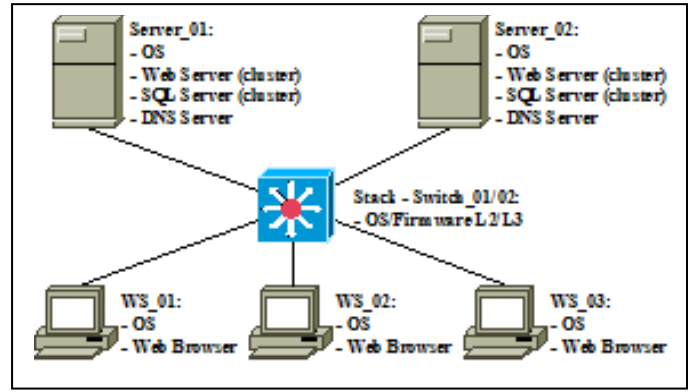

Figure 5. Simple example of multy-layered system model – physical architecture layer

**map_(layer(3), component_(web_server,1),
component_(vserver,1), parameters_([])).**
**map_(layer(3), component_(ss,1),
component_(vserver,1),parameters_([])).**

Generally, the input data for the model construction, i.e. the list of components, the variables Type and the lists Param, must be derived from the technical specifications and subsequently formalized into the set of logical formal operations. We need to state here − the automated transformation techniques (transformation technical specifications and end-user requirements to the logical facts) are beyond the scope of this paper. This problem requires a separate discussion: in the case of complex or non-standard systems, it may not be a routine exercise in practice.

*B. Test requirements*

In our case, the test requirements can be interpreted as input entities of the form of operators for system analysis processes (see Fig. 1). The distributed aspects of requirements can be represented as Prolog fact:

**requirement_(layer_(N), component_(Class_1, Num_1),
component_(Class_2, Num_2),
parameter_([Param])).**

where *component_(Class_1, Num_1)* and *component_(Class_2, Num_2)* is a pair of interacting components on layer *N* (the potential paths of the multi-layered graph); and *[Param]* is a list of technical parameters of component-to-component connections (ports, protocols, modes, etc.).

*Num_1* and *Num_2* can be anonymous variables. The anonymous variable *Num_1* for the term *component_(Class_1, Num_1)* defines all existing components *Class_1* of the system model (the conjunction of components *Class_1*). In turn, the anonymous variable *Num_2* for the term *component_(Class_2, Num_2)* defines at least one component *Class_2* of the system model (the disjunction of components *Class_2*).

In contrast to the facts for component-to-component connections representation, the facts for requirements representation define paths (not edges) but have the same structure. For a simple example of multi-layered system model (see Fig. 2 – Fig. 5), the system requirements are represented as:

**requirement_(layer(4), component_(subscriber,_),
component_(provider,_), parameters_([])).**
**requirement_(layer(3),component_(ss,_),
component_(dns_server,_), parameters_([])).**
**requirement_(layer(2),_,_,_).**
**requirement_(layer(1),_,_,_).**

In this example, the anonymous variable for the term *component_(ss, _)* defines all seven components *ss* (system services/processes) of the system model (see Fig. 3): *component_(ss, _) => component_(ss, 1) ∧ component_(ss, 2) ∧ … ∧ component_(ss, 7)*. And the anonymous variable for the term *component_(dns_server, _)* defines at least one of the two components *dns_server*: *component_(dns_server, _) => component_(dns_server, 1) ∨ component_(dns_server, 2)*.

This form of representation makes the system model useful for error and attack tolerance modeling [27].

*C. Test strategy*

The first step of the test strategy can be achieved by a Prolog predicate such as the following:

**first_step_(N, Type_of_Component) :-
findall(X, object_(layer(N), _, type(X), _), List),
sort(List, Type_of_Component).**

A result of this Prolog query is lists of software-based and hardware-based components:

*['Mozilla Firefox 29.0', 'MySQL Server 5.6'…]*
*['Cisco Catalyst 3750G-24T', 'HP ProBook 450'…]*

As it was sad before, the tests of individual components themselves are usually prepared by vendors and/or independent laboratories and, as a consequence, they are beyond the scope of this work. But this step makes sure that work has been done properly and all components have statutory certificates for inclusion in the project documentation.

It is not necessary for this example, but generally this step of the test strategy can also be used for control of well-posed distribution of hardware resources (specification control) across a complex distributed system.

The second step of the test strategy is based on two main processes of mapping/projection:

1. The mapping of test requirements and interlayer projections to the system model on layer *N*, i.e. horizontal projections (see Fig. 1). The core of this process is the algorithm of finding acyclic paths in graphs [26]:

**path_checking(N, [A1, A2], [[A1, A2] | Path_1],
[[A1, A2] | Path_1]).**

**path_checking(N, [A1, A2], [[Y1, Y2] | Path_1], Path) :-
relation_(N, [X1, X2], [Y1, Y2]),
\+ member([X1, X2], Path_1),
path_checking(N, [A1, A2], [[X1, X2], [Y1, Y2] |
Path_1], Path).**

2. The mapping of paths from layer *N* to layer *N-1*, i.e. vertical or interlayer top-down projections (see Fig. 1), can be divided into three sub steps:

2.1. Mapping the start node of the path:

**path_mapping_1(_, [], []).**

**path_mapping_1(N, [[[A1, A2], [Z1, Z2]] |
Active_Objects_List], List3) :-
findall([A11, A12], map_(layer(N), component_(A1,
A2), component_(A11, A12), _),
Components_List1),
path_mapping_2(X, Z1, Z2, Components_List1,
List2),
append(List2, List1, List3),
path_mapping_1(N, Active_Objects_List, List1).**

2.2. Mapping the end node of the path:

**path_mapping_2(_, _, _, [], []).**

**path_mapping_2 (N, Z1, Z2, [Component1 |
Components_List1], List3) :-
findall([Z11, Z12], map_(layer(N), component_(Z1,
Z2), component_(Z11, Z12), _),
Components_List2),
pair_gen(Component1, Components_List2 , List2),
append(List2, List1, List3),
path_mapping_2(N, Z1, Z2, Components_List1,
List1).**

2.3. Generation pairs of components (determination the potential paths) for layer *N-1*:

**pair_gen(_, [], []).**

**pair_gen(Component1, [Component2 |
Components_List2], [[Component1,
Component2] | List2]) :-
pair_gen(Component1, Components_List2, List2).**

The requirements-coverage strategy application to the simple example of multi-layered system model shows a surprisingly large number of tests required to fully cover even this very simple model – see Tab. 1.

TABLE I.    APPLICATION OF THE REQUIREMENTS-COVERAGE STRATEGY

| Model layers | Test templates | | |
|---|---|---|---|
| | Individual components | Distributed aspect | |
| | $T_{comp}$ | $T_{dist}$ | Symbolic description |
| Functional | 2 | 1 | [subscriber, 1] ↔ [provider, 1] |
| System | 14 | 3 | [web_client,_] ↔ [web_server,1] ↔ [sql_server,1] |
| | | 12 | [ss,_] ↔ [dns_server,_] |
| Logical | 6 | 6 | [vws,_] ↔ [vlan,0] ↔ [vserver,_] |
| | | 1 | [vserver,1] ↔ [vlan,0] ↔ [vserver,2] |
| Physical | 7 | 6 | all component-to-component connections |

## V.  NEXT STEPS

The complex testing of distributed systems (as every testing activity) is based on two basic notations [2]:

- test philosophy (or test strategy) that establishes "what" should be tested;
- test methodology that establishes "how" it should be done.

In this work we have determined the test templates generation strategy. On the other hand, The Integrated Test Methodology for Distributed Systems [28] includes the following general procedures: (1) conformance testing; (2) interoperability testing; (3) functional testing; and (4) performance testing (in spite of the year of publication, this methodology is not contrary to the most recent work, include testing of cloud computing [29]).

The conjunction of the requirements-coverage strategy (the test philosophy) and the test methodology procedures can be represented as a testing grid – see Fig. 6. Intersection points of this grid determine test cases which cover the most important goals of distributed systems [1]: (1) openness; (2) accessibility; and (3) transparency (scalability of distributed systems is beyond the scope of this work). In turn, these goals are supported by their performance characteristics.

The final step must be the processing of results (according to the requirements of Standard IEEE Std 829TM-2008 (Revision of IEEE Std 829-1998) [30]) for inclusion in the project documentation.

## VI.  CONCLUSION

Deployment of distributed systems sets high requirements for procedures, tools and approaches for complex testing of these systems. In this work we determined the formal model for test generation mission on the basis of the component based approach and the concept of layered networks. The model is the four-layered 3D graph, derived from the system technical specifications, which covers all layers of OSI Reference Model and, as a consequence, both software-based and network-based aspects of distributed systems.

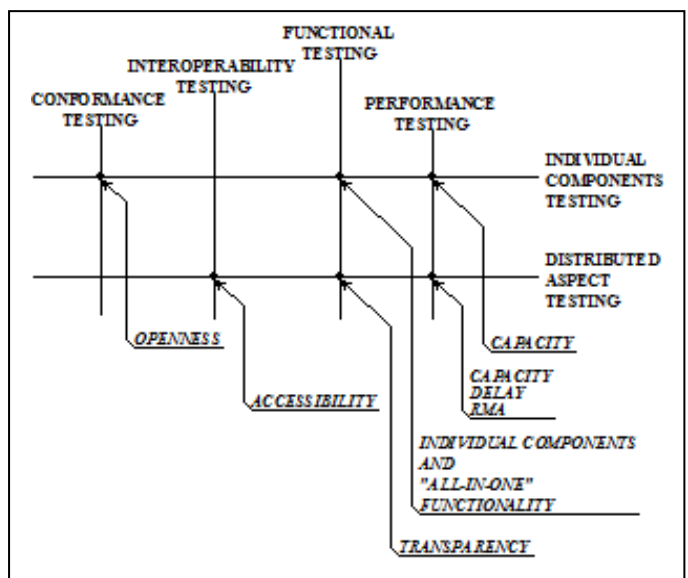

Figure 6.  Testing grid

Based on this model, we described the test templates generation strategy, which covers: (1) individual components; (2) every interaction from the end-user requirements on functional, system, logical and physical architectural layers; and, as a complement, checks the internal consistency of the system technical specifications with respect to the end-user requirements.

Next, we introduced a Prolog-based approach to representing the formal multi-layer model and the requirements-coverage strategy.

The most important conditions that should be fulfilled to make building a distributed system worth the effort [1]: (1) openness; (2) accessibility; and (3) transparency. The requirements-coverage strategy completely covers the first two goals and partially the third one with the exception of the failure transparency in the case of imperfect sensing and switching components [31]. As a consequence, future work will focus on design of a test philosophy for determination of the necessary set of test cases for reliability testing of complex distributed systems (it would be better to talk of a necessary and sufficient set, but unfortunately in our case a sufficient condition is theoretically unreachable [32]).

ACKNOWLEDGMENT

This research has been performed within the scientific activities at the Department of Telecommunication Engineering of the Czech Technical University in Prague, Faculty of Electrical Engineering.

REFERENCES

[1]  A. S. Tanenbaum and M. v. Steen, Distributed Systems: Principles and Paradigms, 3rd ed., Prentice Hall Press, 2013.

[2]  R. W. Buchanan, The art of testing network systems, John Wiley & Sons, 1996.

[3] D. M. Buede, The Engineering Design of Systems: Models and Methods, 2nd ed., Wiley Publishing, 2009.

[4] S. R. Dalal, A. Jain, N. Karunanithi, J. M. Leaton, C. M. Lott, G. C. Patton and B. M. Horowitz, "Model-based testing in practice," in *Software Engineering, 1999. Proceedings of the 1999 International Conference on*, 1999.

[5] R. M. Hierons, K. Bogdanov, J. P. Bowen, R. Cleaveland, J. Derrick, J. Dick, M. Gheorghe, M. Harman, K. Kapoor, P. Krause, G. Luttgen, A. J. H. Simons, S. Vilkomir, M. R. Woodward and H. Zedan, "Using formal specifications to support testing," *ACM Comput. Surv.,* vol. 41, pp. 9:1-9:76, February 2009.

[6] J. Woodcock, P. G. Larsen, J. Bicarregui and J. Fitzgerald, "Formal methods: Practice and experience," *ACM Comput. Surv.,* vol. 41, pp. 19:1-19:36, October 2009.

[7] J. J. Marciniak, Ed., Encyclopedia on Software Engineering, Wiley, 2001.

[8] P. Stocks and D. Carrington, "A framework for specification-based testing," *Software Engineering, IEEE Transactions on,* vol. 22, pp. 777-793, 1996.

[9] J. Offutt and A. Abdurazik, "Generating Tests from UML Specifications," in *The Unified Modeling Language*, R. France and B. Rumpe, Eds., Springer Berlin Heidelberg, 1999, pp. 416-429.

[10] J. Hartmann, C. Imoberdorf and M. Meisinger, "UML-Based Integration Testing," *SIGSOFT Softw. Eng. Notes,* vol. 25, pp. 60-70, August 2000.

[11] G. Bernot, M.-C. Gaudel and B. Marre, "Software testing based on formal specifications: a theory and a tool," *Software Engineering Journal,* vol. 6, pp. 387-405, 1991.

[12] J. Dick and A. Faivre, "Automating the Generation and Sequencing of Test Cases from Model-Based Specifications," in *Proceedings of the First International Symposium of Formal Methods Europe on Industrial-Strength Formal Methods*, 1993.

[13] M. R. Donat, "Automating formal specification-based testing," in *TAPSOFT '97: Theory and Practice of Software Development, 7th International Joint conference CAAP/FASE*, 1997.

[14] Hyoung Seok Hong, Sung-Deok Cha, Insup Lee, O. Sokolsky and H. Ural, "Data flow testing as model checking," in *Software Engineering, 2003. Proceedings. 25th International Conference on*, 2003.

[15] Shaoying Liu and Wuwei Shen, "A formal approach to testing programs in practice," in *Systems and Informatics (ICSAI), 2012 International Conference on*, 2012.

[16] ISO/IEC, *ITU-T Rec. X.200 - ISO/IEC 7498:1994 Information technology - Open Systems Interconnection - Basic Reference Model,* 1994.

[17] J. Liu and E. A. Lee, "A component-based approach to modeling and simulating mixed-signal and hybrid systems," *ACM Trans. Model. Comput. Simul.,* vol. 12, pp. 343-368, October 2002.

[18] M. Torngren, DeJiu Chen and I. Crnkovic, "Component-based vs. model-based development: a comparison in the context of vehicular embedded systems," in *Software Engineering and Advanced Applications, 2005. 31st EUROMICRO Conference on*, 2005.

[19] M. Kurant and P. Thiran, "Layered Complex Networks," *Phys. Rev. Lett.,* vol. 96, no. 13, April 2006.

[20] J. D. McCabe, Network Analysis, Architecture, and Design, 3rd ed., Morgan Kaufmann Publishers, 2007.

[21] ISO/IEC, *ITU-T Rec. X.901-904 - ISO/IEC 10746 Information technology - The Reference Model of Open Distributed Processing (RM-ODP),* 1998.

[22] F. Brazier, B. D. Keplicz, N. R. Jennings and J. Treur, "Formal Specification of Multi-Agent Systems: a Real-World Case," in *First International Conference on Multi-Agent Systems, ICMAS-95*, 1995.

[23] B. Beizer, Black-box testing: techniques for functional testing of software and systems, John Wiley & Sons, 1995.

[24] R. Diestel, Graph Theory, 4th ed., Springer, 2010.

[25] D. K. Pradhan, Ed., Fault-tolerant computer system design, Prentice-Hall, 1996.

[26] I. Bratko, Prolog: Programming for Artificial Intelligence, 4th ed., Addison-Wesley Longman, Inc., 2012.

[27] M. Kurant, P. Thiran and P. Hagmann, "Error and Attack Tolerance of Layered Complex Networks," *Phys. Rev. E,* vol. 76, no. 2, August 2007.

[28] J. Grabowski and T. Walter, "Towards an Integrated Test Methodology for Advanced Communication Systems," in *Proceedings of the '16th International Conference and Exposition on Testing Computer Software (TCS'99)*, 1999.

[29] K. Blokland, J. Mengerink and M. Pol, Testing Cloud Services, 1st ed., Rocky Nook Inc., 2013.

[30] IEEE, *IEEE Std 829TM-2008 - IEEE Standard for Software and System Test Documentation,* 2008.

[31] M. Modarres, M. Kaminskiy and V. Krivtsov, Reliability Engineering And Risk Analysis: A Practical Guide, 2nd ed., CRC Press, 2010.

[32] N. G. Leveson, Safeware: system safety and computers, ACM, 1995.